\documentclass[a4paper,11pt]{article}
\pdfoutput=1 % if your are submitting a pdflatex (i.e. if you have
\usepackage{jcappub} % for details on the use of the package, please
\usepackage{color}
\usepackage[T1]{fontenc} % if needed
\usepackage[modulo]{lineno}

\title{\texttt{simsurvey}: Estimating Transient Discovery Rates for the Zwicky Transient Facility}

\author[a,1]{Ulrich~Feindt,\note{Corresponding author.}}
\author[b]{Jakob~Nordin,}
\author[c]{Mickael~Rigault,}
\author[b]{Val\'{e}ry~Brinnel,}
\author[a]{Suhail~Dhawan,}
\author[a]{Ariel~Goobar,}
\author[b,d]{and Marek~Kowalski}

\affiliation[a]{The Oskar Klein Centre, Department of Physics, Stockholm University, AlbaNova, SE-106~91 Stockholm, Sweden}
\affiliation[b]{Institute of Physics, Humboldt-Universit\"at zu Berlin, Newtonstr. 15, 124 89 Berlin, Germany}
\affiliation[c]{Universit\'e Clermont Auvergne, CNRS/IN2P3, Laboratoire de Physique de Clermont, F-63000 Clermont-Ferrand, France}
\affiliation[d]{Deutsches Elektronensynchrotron, Platanenallee 6, D-15738, Zeuthen, Germany}

\emailAdd{ulrich.feindt@fysik.su.se}
\emailAdd{jnordin@physik.hu-berlin.de}
\emailAdd{m.rigault@ipnl.in2p3.fr}
\emailAdd{vbrinell@physik.hu-berlin.de}
\emailAdd{suhail.dhawan@fysik.su.se}
\emailAdd{ariel@fysik.su.se}
\emailAdd{marek.kowalski@desy.de}

\abstract{When planning a survey for astronomical transients, many
  factors such as cadence, filter choice, sky coverage, and depth of
  observations need to be balanced in order to optimize the scientific
  gain of the survey.  Here we present a software package called
  \texttt{simsurvey} for simulating the supernova lightcurves that are
  expected based on a survey strategy, which can then be used to
  determine the potential for discoveries of each strategy in
  question. The code is set up in a modular fashion that allows easy
  modification of small details of the survey and enables the user to
  adapt it to any survey design and transient template that they wish
  to use in planning their survey. As an example of its utility, we use
  \texttt{simsurvey} to simulate the lightcurve of several types of
  supernovae that the recently started Zwicky Transient Facility (ZTF)
  is expected to find and compare the results to the discoveries made
  during its early operations. We conclude that ZTF will find thousands
  of bright supernovae per year, of which about 10 could potentially
  be found with two days of explosion. Over the course of three years
  the survey will obtain lightcurves of about 1800 type Ia supernovae
  with $z < 0.1$ that can be used as distance indicators in cosmology
  if they are spectroscopically classified using additional
  telescopes. In a comparison to detections from the ZTF
  public survey, we found good agreement with the numbers of
  detections expected from the simulations.}

\begin{document}
\maketitle
\flushbottom

\section{Introduction}
\label{sec:intro}

Over the past decades the search for transient astronomical events,
such as supernovae, has evolved from the occasional serendipitous
discovery to systematic searches using dedicated telescopes designed
to find transients. Such surveys, including the Supernova Legacy
Survey (SNLS; \cite{2006A&A...447...31A}), the Sloan Digital Sky
Survey II (SDSS-II, \cite{2008AJ....135..338F}), Palomar QUEST
\cite{2008AN....329..263D}, Skymapper \cite{2007PASA...24....1K},
PanSTARRS \cite{2010SPIE.7733E..0EK}, the All-Sky Automated Survey for
Supernovae (ASAS-SN; \cite{2014AAS...22323603S}), the Asteroid
Terrestrial-impact Last Alert System (ATLAS;
\cite{2018PASP..130f4505T}), the Dark Energy Survey (DES;
\cite{2005astro.ph.10346T}), and the (intermediate) Palomar Transient Factory
(PTF/iPTF; \cite{2009PASP..121.1395L}), have systematically discovered
thousands of supernovae.

One of the latest transient surveys is the Zwicky Transient Facility
(ZTF, \cite{ztf-graham, 1538-3873-131-995-018002}) using the Palomar
48-inch Schmidt telescope. The same telescope has previously been used
for PTF/iPTF and has now been equipped with a new camera that allows
full usage of its 47~sq.~deg. field of view and shorter
exposures. With this improvement in survey speed, ZTF can practically
cover the whole observable sky in a single night (or observe the same
location repeatedly), reaching $\sim$~20.5~mag in 30 second
exposures. This makes it an excellent survey for a great range of
transient studies because it can survey a large nearby volume, in
which spectroscopy is easier to obtain, by covering the sky
rapidly. ZTF can find supernovae within hours of explosion, with the 
potential for discovering new types of events that evolve much faster
and it can also collect a large and well-controlled sample of
supernovae for studies of their rates and population properties. ZTF can 
also, in case of type Ia supernovae, provide accurate measurements
of distances in the local universe that can be used for cosmological
inference.

In order to make the best use of an instrument like the ZTF camera,
careful planning of the survey strategy is required, especially when
varied scientific interests need to be balanced. For previous surveys
the strategy has been optimized using Monte Carlo methods that
generate lightcurves of transient that will be obtained by the survey
based on the survey strategy, e.g.\ using the package \texttt{SNANA}
\cite{2009PASP..121.1028K} for DES
\cite{2012ApJ...753..152B,2017ApJ...837...57D} and the Wide-Field InfraRed
Survey Telescope (WFIRST, \cite{2017arXiv170201747H}). Additionally
this technique can be used to determine the survey efficiencies and biases as has
been done for DES \cite{2015AJ....150..172K} and the joint analysis of
SDSS and SNLS SNe~Ia data \cite{2014ApJ...793...16M}.

In this paper we present a Python package, \texttt{simsurvey}, that
can be used for this type of analysis.  Based on these lightcurves we
can then define metrics for how well the survey works for a given
science case, e.g.\ how many SNe can be found shortly after explosion
or for how many SNe Ia will we obtain the lightcurve coverage required
for cosmology.  We will use ZTF as an example of a survey, for which
we used the package to optimize the strategy. However, we would like
to point out that this package can be used for any survey and is not
restricted to ZTF. All parameters that differ between surveys,
e.g.\ CCD layout or field grid definition, can be adjusted easily.

In section \ref{sec:simsurvey} we present the \texttt{simsurvey}
Python package that is used to simulate the lightcurves that we expect
from ZTF. We outline the simulation in \ref{sec:sim-ztf} and show the
results in section \ref{sec:results}. The scientific implications of
the simulations are discussed in section \ref{sec:discussion} and
compared to the first discoveries by ZTF in section
\ref{sec:comp-ztf}. Finally we summarize our findings in section
\ref{sec:concl}.

\section{Survey simulation software \texttt{simsurvey}}
\label{sec:simsurvey}

The lightcurves are simulated using Python code mostly from the
packages \texttt{sncosmo}~\cite{2016ascl.soft11017B} and
\texttt{astropy}~\cite{2013A&A...558A..33A,2018AJ....156..123T}. This
code has been released on PyPI as \texttt{simsurvey}
\cite{ulrich_feindt_2019_2554337} and will be described 
below\footnote{Source code and a more extensive documentation are 
available at \url{https://github.com/ufeindt/simsurvey}}. When
simulating the lightcurves of a whole survey, the following input will
be required:
\begin{itemize}
\item A survey schedule containing the time, pointing and filter of
  each exposure. Additionally sky brightness and zero points can be
  provided for each observation in the schedule, e.g.\ based on a
  weather model, or otherwise these parameters can be set to a default
  value.
\item A transient model, i.e.\ a time-series of spectral energy
  distributions (SEDs) from which the photometry of the transients can
  be calculated. The SEDs of the transient need to be provided over a
  sufficient wavelength range such that all filters used in the survey
  schedule fall into it for the whole redshift range of the
  simulation.
\item A function that allows us to sample the transient model
  parameters, e.g.\ peak magnitude, lightcurve width or host galaxy
  extinction for each transient, from a distribution modeling the
  transient population.
\end{itemize}

As a first step, transients are placed at redshifts sampled from a
distribution based on the volumetric rate as a function of redshift up
to a redshift cutoff. The total number of transients is based on the
integrated rate as well as the time span of the survey and the solid
angle it covers but it can also be fixed to any arbitrary value instead of
using the absolute rate to determine it. In this case the relative
redshift distribution will remain the same. For each transient the
code draws the coordinates on the sky from a uniform distribution on
the unit sphere that can be limited in right ascension and
declination, and should be chosen to cover the whole survey
footprint. The Milky Way can be excluded using a simple cut in
Galactic latitude $b$, but this is not necessary in order to exclude
transients behind the Galactic plane because the code will determine
the extinction by Galactic dust for all transients based on the maps
from \cite{1998ApJ...500..525S} and apply extinction to the transient
model. Thus while lightcurves for transients behind the Milky Way
would be simulated if the survey schedule points in that direction,
they will generally be below the detection threshold due to
extinction. Lastly a random Julian date for the $t_0$ parameter (time
of peak or time of explosion) of the model is drawn from a uniform
distribution based on the time span of the survey.

Based on the redshifts and the transient template the other parameters
for the lightcurve simulation are then determined. For several
built-in templates of \texttt{sncosmo}, simple distributions of the
model parameter reflecting the transient population have been built
into \texttt{simsurvey} but both the template and the distribution can
easily be replaced by the user. Instead of using one of the built-in 
templates, a custom transient template can be defined using \texttt{sncosmo}. 
These templates can, for instance, be based on SEDs obtained from simulations 
of the merger of two neutron stars. Alternatively analytic functions can also 
be used to define the model. An example of this is using Planck's law of 
black-body radiation along with formulae for the evolution of temperature and 
radius of the photosphere of a transient. Therefore, \texttt{simsurvey} can be 
used to simulate both well-studied SNe and hypothetical transients that are 
significantly rarer.

Distributions for all transient model parameters except the redshift
and $t_0$ can be defined by a single Python function. The list of model
parameters usually includes the flux normalization of the model --
based on the transient's absolute magnitude distribution and its
distance modulus -- and model-specific parameters, e.g.\ the
SALT2~\cite{2007A&A...466...11G} stretch and color parameters $x_{1}$
and $c$. Additionally, \texttt{sncosmo} models can be extended by
propagation effects such as extinction by dust in the host galaxy 
(modeled by a wavelength-dependent extinction law). These effects add 
further parameters to the model, e.g.\ $E(B-V)$ and $R_V$.

Using this set of transient model parameters the code can then
simulate lightcurves based on the survey schedule. First a list of
survey fields, in which a transient is located, is generated for each
transient. For this, the shape of the survey's field of view must be
provided. In the simplest case this can be the projection of
rectangular field of view onto the sky (defined by its width and
height in degrees) but additionally the code can match the transients
to individual CCD chips of a camera if its layout is provided.
\texttt{simsurvey} can therefore also be used for surveys with a field
of view that is more complicated than a simple rectangle, e.g.\ LSST,
and also account for losses due to the gaps between the CCD
chips. Once the fields in which each transient is located have been
determined, a list of observations that include its coordinates is
generated. If the survey strategy does not use a fixed grid of fields
for observations, the pointings can also be given by just their
coordinates, which will then be matched with the transients
individually.

The observations list is further restricted by the minimum and
maximum times relative to $t_0$ for which the template is defined
(multiplied by $(1+z)$ to account for time dilation) in order to limit
the lightcurve to the relevant parts if the survey covers a much
longer time scale. The time frame is extended by two weeks prior to
the definition range of the model, for which the flux is set to a
random value based on the sky brightness. This allows us to assess
whether we would have pre-explosion limits for the transient in
question. This procedure does not account for SN precursor events 
that may occur for some SN types, especially type IIn, more than two 
weeks prior to explosion \cite{2014ApJ...789..104O}. However, such a 
precursor could be added to an existing SN model using 
\texttt{sncosmo}.\footnote{\texttt{simsurvey} contains code, specifically 
the \texttt{CompoundSource} class, with which such a model could easily be 
constructed by combining two or more templates and allowing a varying time 
difference between them, see the online documentation for more information.}
For all other pointings to the transient, the flux is
calculated based on the template and the passbands used by the survey
and then perturbed by noise terms based on the sky brightness,
the Poisson noise expected for the flux of the transient, and the gain
of the instrument. 

The lightcurves are then filtered using selection
criteria that correspond to the discovery process of transient
surveys. The default transient filter requires two
$5\sigma$-detections in the same night for a transient to be included
in the final output. The requirement of two detections simulates the
common asteroid rejection method since the pointings are generally
scheduled at least half an hour apart and most asteroids will move
sufficiently in that time and thus not be detected as transients at
the same coordinates. Lastly useful statistics of the lightcurves
will be calculated at the same time, e.g.\ the time of discovery and
the the interval between discovery and the last non-detection,
i.e.\ an observation of the transients locations without a detection.

\section{Simulation of the ZTF survey}
\label{sec:sim-ztf}

\subsection{Survey schedule}
\label{sec:sim-ztf-sched}

The observing time for the Palomar 48-inch telescope (P48) within ZTF
has been allocated to different programs, 40\% of the time each go to
the public survey and the ZTF partner institutes while the remaining
20\% can be used by Caltech staff. In this paper we will disregard the
Caltech time and only focus on the extra-galactic parts of the other
two programs. A full discussion of the ZTF survey strategy can be
found in \cite{ztf-bellm-scheduler}.

The public survey observes the whole visible sky in $g$- and
$r$-band at a three-day cadence. Observations of the same field in
both bands are scheduled for the same night with sufficient time
in between them to distinguish extra-galactic transients from
asteroids.

The ZTF partnership time is being used for various smaller
extra-galactic surveys\footnote{Additionally it will include surveys
  of the Galactic Plane and the Solar System, which will not be
  discussed here.}. Most of the time is used for a high-cadence survey
of $\sim1600$~sq.~deg.\ with up to six $g$-band observations per night
that are separated by at least 30 minutes. This leads to a large
number of early transient detections, for some of which we will be
able to constrain the time of explosion very well. Furthermore a
survey of $\sim6700$~sq.~deg.\ with a four-day cadence in the $i$-band
will be carried out, adding a third filter to many lightcurves from
the public survey, which will greatly improve cosmological distance
estimates from SNe Ia. Lastly, time is also used for
target-of-opportunity (ToO) observations for multi-messenger
astronomy, following up on gravitational wave, neutrino or gamma-ray
burst triggers.

Simulating full survey operations (including e.g.\ slew times and
filter changes) would be beyond the scope of this study. Instead we
select fields (from a grid of 879 fields covering the whole sky) that
are observable during a given night and schedule them in a way that
corresponds to the cadence of the subprogram without accounting for
when exactly the field is at a low airmass during the
night. Additionally we did not account for changes in the observing
conditions, e.g.\ due to differences in airmass and sky brightness,
but instead set the $5\sigma$-depth of all observations to
20.5~mag. For the simulation we have chosen a period of one year from
February 1st, 2018 to January 31st, 2019. The relevant programs of the
survey were simulated using the following scheduling:

\begin{itemize}
\item \textbf{public:} For every third night of the survey we selected all fields
  that are at an airmass < 2 for at least an hour during that night and that have
  low Milky Way extinction ($E(B-V) \leq 0.2$). A third of those fields are then
  scheduled for $g$- and $r$-band observation in the selected night and the
  remaining fields are observed in the two nights after it.
\item \textbf{high-cadence:} This program was limited to 34 of the
  public survey fields (corresponding to $\sim1600$~sq.~deg.)
  described above that are observable in at least 95\% of the nights
  throughout an eight-month period from March to October. In addition,
  the fields were selected such that they are no closer than
  $50^{\circ}$ to the full moon. (This requirement is lowered based on
  the phase of the moon.)
\item \textbf{$i$-band:} The area covered by the $i$-band program is too
  large for a single set of fields as used for the high-cadence
  program. Instead fields were selected such that they can be observed
  for at least 3.5 months in the period between March and November. For
  each month we selected 143 fields($\sim6700$~sq.~deg.) to be
  observed while making sure that individual fields are observed for at
  least 3 months. A subset of the fields are observable for longer
  than that, including the 34 fields for the high-cadence program. In
  total this survey covers $\sim12000$~sq.~deg..
\end{itemize}

\begin{figure}[tp]
  \centering
  \includegraphics[width=0.49\textwidth]{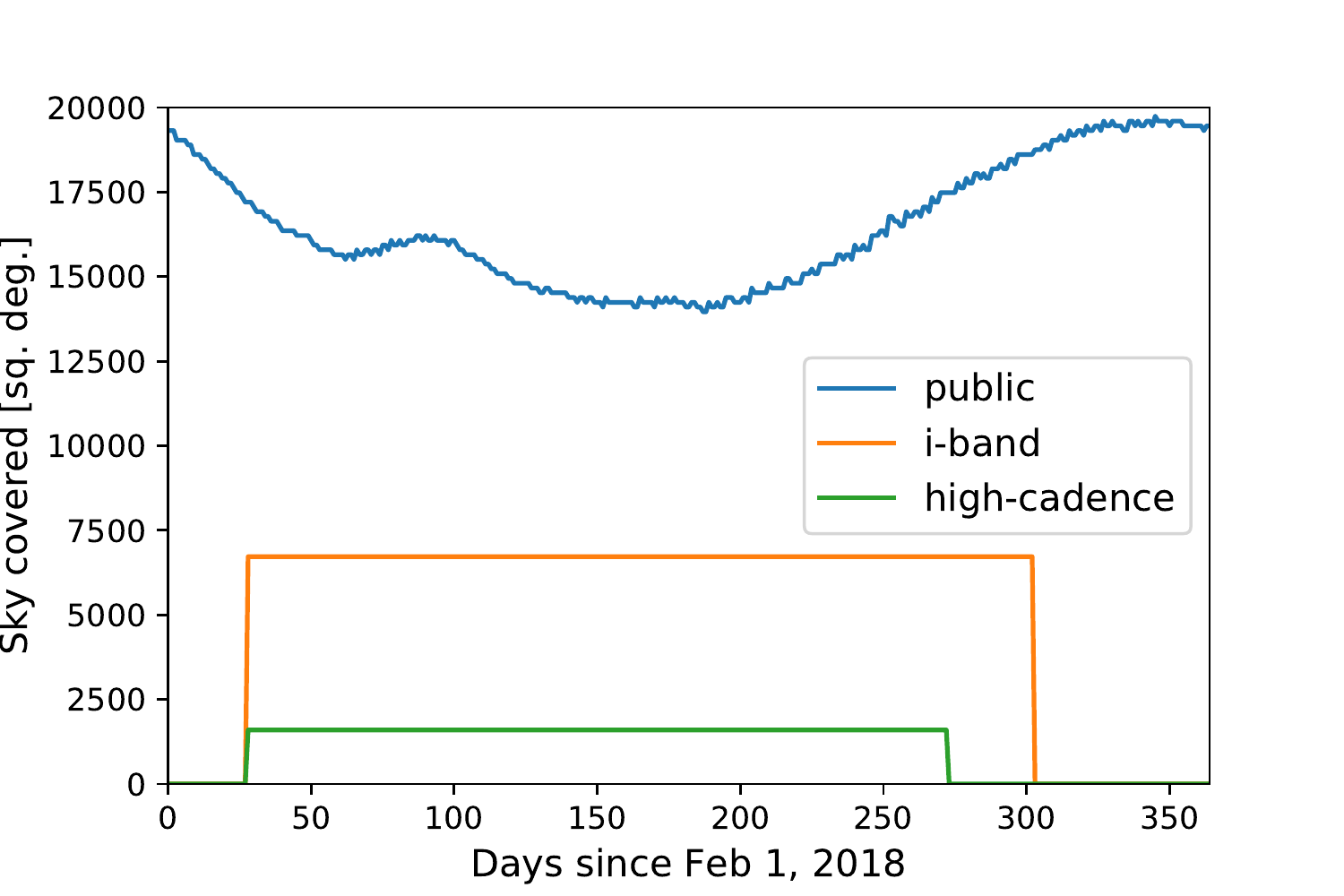}
  \includegraphics[width=0.49\textwidth]{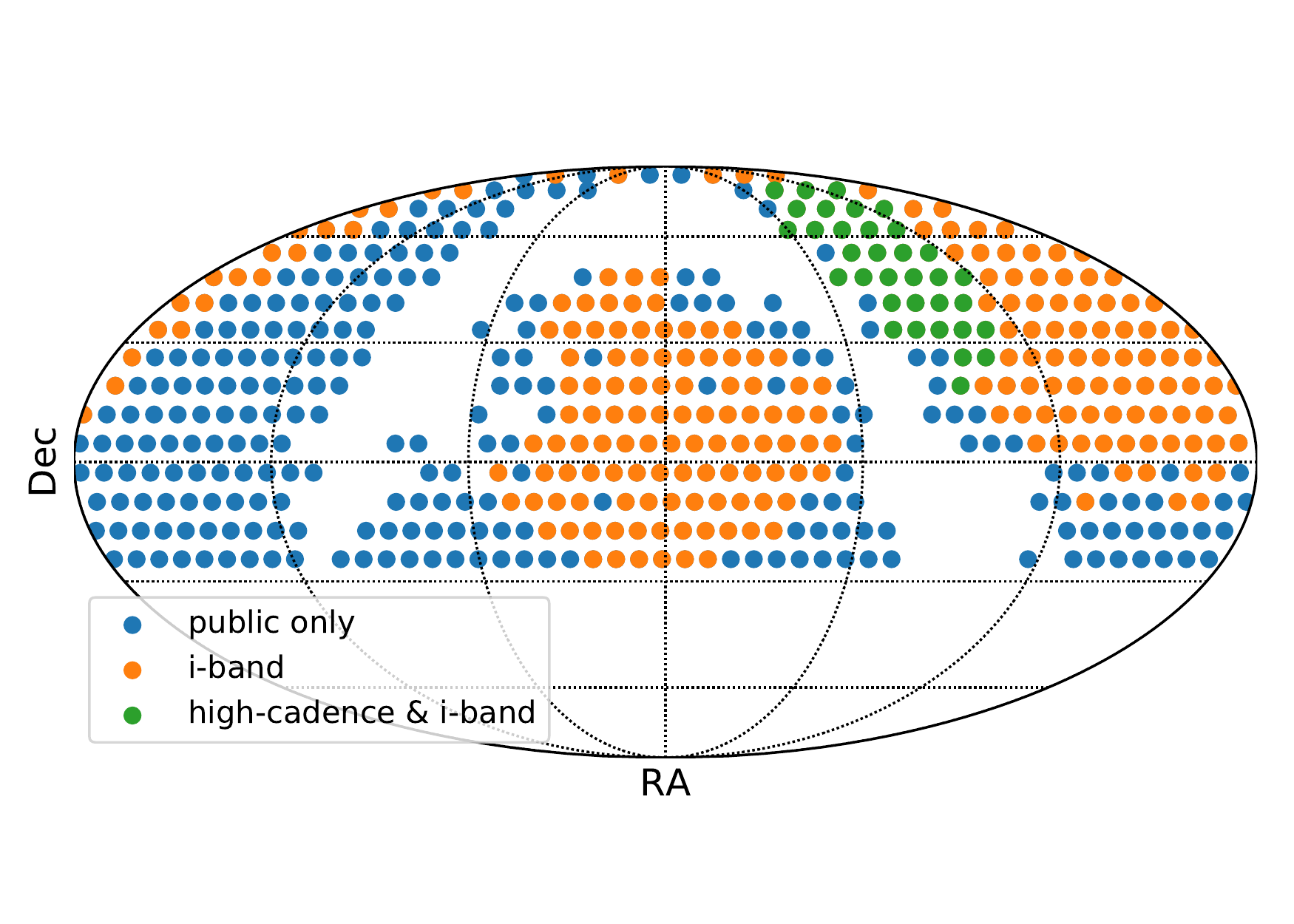}
  \caption{\emph{Left:} Solid angle in square degrees monitored by the
    programs of the survey throughout the year. (The whole sky has a
    solid angle of $\sim41000$~sq.~deg.) \emph{Right:} Skymap of the
    pointings of each program. Note that the marker shape is not the
    shape of the ZTF field of view and that the true shapes of the ZTF
    field of view overlap by 13\% on average. All shown pointings are
    in the public survey and all fields in the high-cadence survey are
    covered by the $i$-band survey as well.}
  \label{fig:programs-sky}
\end{figure}

\noindent The field selection for each survey and the area covered by
it are summarized in Fig.~\ref{fig:programs-sky}. Note that these
survey schedules are not using exactly the same field selection as the
actual survey (see \cite{ztf-bellm-scheduler}), e.g.\ the public
extra-galactic survey did not have a cutoff in Milky way extinction but
rather in Galactic latitude $b$, and the high-cadence survey ran beyond
the end of October where most fields were setting at that time and became
more difficult to observe multiple times per night. The simplified
surveys used here are, however, still sufficiently close to the actual
survey to yield useful estimates.

\subsection{Lightcurve generation}
\label{sec:lc-gen}

\begin{table}[tp]
	\caption{Supernova models used for the simulation and their
          main simulation parameters. $^\dagger$In addition to the
          intrinsic scatter of SN~Ia peak magnitudes, the Tripp
          relations \cite{1998A&A...331..815T} were used to simulate a
          realistic population. $^\ddagger$To avoid simulating a large
          number of unrealistically bright SNe~IIn, the Gaussian
          distribution of peak magnitudes was truncated at $1\sigma$
          on the brighter-than-average side.}
  \vspace{0.2cm}
	\label{tab:sn-templates}     
	\centering                
	\begin{tabular}{ccccc}
    %& \texttt{sncosmo} & Rate & \\
    SN type & \texttt{sncosmo} template & Rate [Mpc$^{-3}$ yr$^{-1}$] & $M_{B}$ (peak) & $\sigma_M$ \\
    %\vspace{0.1cm}
    \hline
    %\vspace{0.1cm}
    Ia & \texttt{salt2} & $3\times10^{-5}$ & $-19.3$ & $0.1^\dagger$\\
    Ib/c & \texttt{nugent-sn1bc} & $2.25\times10^{-5}$ & $-17.5$ & $1.2$\\
    IIn & \texttt{nugent-sn2n} & $7.5\times10^{-6}$ & $-18.5$ & $1.4^\ddagger$\\
    IIP & \texttt{nugent-sn2p} & $1.2\times10^{-4}$ & $-16.75$ & $1$\\
	\end{tabular}
\end{table}

For the lightcurve simulations we focused on type Ia supernovae and
common types of core-collapse SNe, specifically types Ib/c, IIn and
IIP. The templates were chosen from the sets of built-in model of
\texttt{sncosmo} and are summarized in
Table~\ref{tab:sn-templates}. The SN models were scaled to peak $B$-band
magnitudes based roughly on the distribution reported in
\cite{2014AJ....147..118R}. In addition extinction by dust in the host
galaxy was added to the models for core-collapse supernovae. For this
we used the extinction law of Cardelli, Clayton \&
Mathis~\cite{1989ApJ...345..245C} with a relatively steep slope
$(R_V=2)$ and $E(B-V)$ drawn from an exponential distribution with a
rate $\lambda=0.11$ (as reported by \cite{2018A&A...615A..45S}). Since
the model used for SNe Ia (SALT2, \cite{2007A&A...466...11G,
  2014A&A...568A..22B}) uses an effective color term that includes
host galaxy extinction, we did not include this effect again. Instead
values for the model parameters $x_1$ (``stretch'') and $c$
(``color'') were drawn from Gaussian distributions centered around 0
with a width $\sigma$ of 1 and 0.1, respectively. Based on these
values the peak magnitude of the SN~Ia is adjusted according to the
Tripp relations \cite{1998A&A...331..815T} that correlate the peak
brightness with lightcurve slope and color and need to be corrected
when using SNe~Ia as distance indicator. The coefficients were chosen
to be close to the usual results of cosmological analyses, e.g.\
\cite{2014A&A...568A..22B}, $\alpha=0.13$ for the stretch and
$\beta=3$ for the color.

The transients were distributed in redshift assuming a constant
volumetric rate.  This assumption is valid because the SNe detected by
ZTF will mostly be near-by.  At the farthest, some SNe Ia may be found
out to $z=0.2$ but the bulk is around $z=0.1$. The coordinates for the
transients are drawn from a uniform distribution (by solid angle) down
to declinations $\delta=-30^{\circ}$, which is the farthest south that
the schedule covers. The Galactic Plane is not explicitly excluded in
the transient placement but an extinction term based on dust in the
Milky Way is added to the model and therefore transients behind the
Galactic Plane (as described in section~\ref{sec:simsurvey}) are less
likely to be observed. The total number of simulated transients was
calculated based on the redshift-integrated volumetric rate multiplied
by the fraction of the sky on which the transients are simulated (75\%
in this case) and the time range for the $t_0$ parameter of the model,
i.e.\ either time of explosion or of peak brightness, which is chosen
to begin one month before the schedule and end one month after it in
order to account for cases, for which only the rising or declining
part of the lightcurve is observed. The rates are listed in
Table~\ref{tab:sn-templates}. For SNe Ia we used the value from
\cite{2011MNRAS.412.1473L} and for the core-collapse SN rate the one
from \cite{2007MNRAS.377.1229M}. We note that the value for the latter
($1.5\times10^{-4}~\mathrm{Mpc}^{-3}\mathrm{yr}^{-1}$) is larger than
the values reported in \cite{2011MNRAS.412.1473L} but still with the
typical uncertainties for these measurement\footnote{Note furthermore
  that the numbers presented here are all proportional to the assumed
  rates and thus can easily be scaled to other estimates}. The
core-collapse SN rate was split the following way: 80\% type IIP (also
including the type IIL rate), 15\% type Ib/c, and 5\% type IIn. This
roughly corresponds to the relative rates of these subtypes according
to \cite{2011MNRAS.412.1441L}.

The simulation was run 100 times for each transient type and a
one-year survey plan in order to build up larger statistics. We did
not use any model to predict the sky brightness but just fixed its
contribution to the flux uncertainty at a value corresponding to a
$5\sigma$-depth of 20.5~mag instead.  At first we assumed perfect
observing conditions for every night when generating the lightcurves
but then certain epochs are removed from them based on the iPTF
pointing history for 2016, therefore also accounting for losses of
observation time that are not due to weather. For simplicity, we only
removed full nights and did not attempt to create a schedule that
accommodate for weather losses in any way, e.g.\ by rescheduling
observations that were missed. All results have been normalized to the
expectation for one year of ZTF operations as described in
section~\ref{sec:sim-ztf-sched}. The script required to rerun the 
simulations can be found at \url{https://github.com/ufeindt/simsurvey-paper-scripts}.

\section{Results}
\label{sec:results}

\subsection{Detection of young transients}
\label{sec:results-young-tr}

\begin{figure}[tp]
  \centering
  %[Plots of discovery phase/time for Ia's and CCs (2 panels)]
  \includegraphics[width=0.49\textwidth]{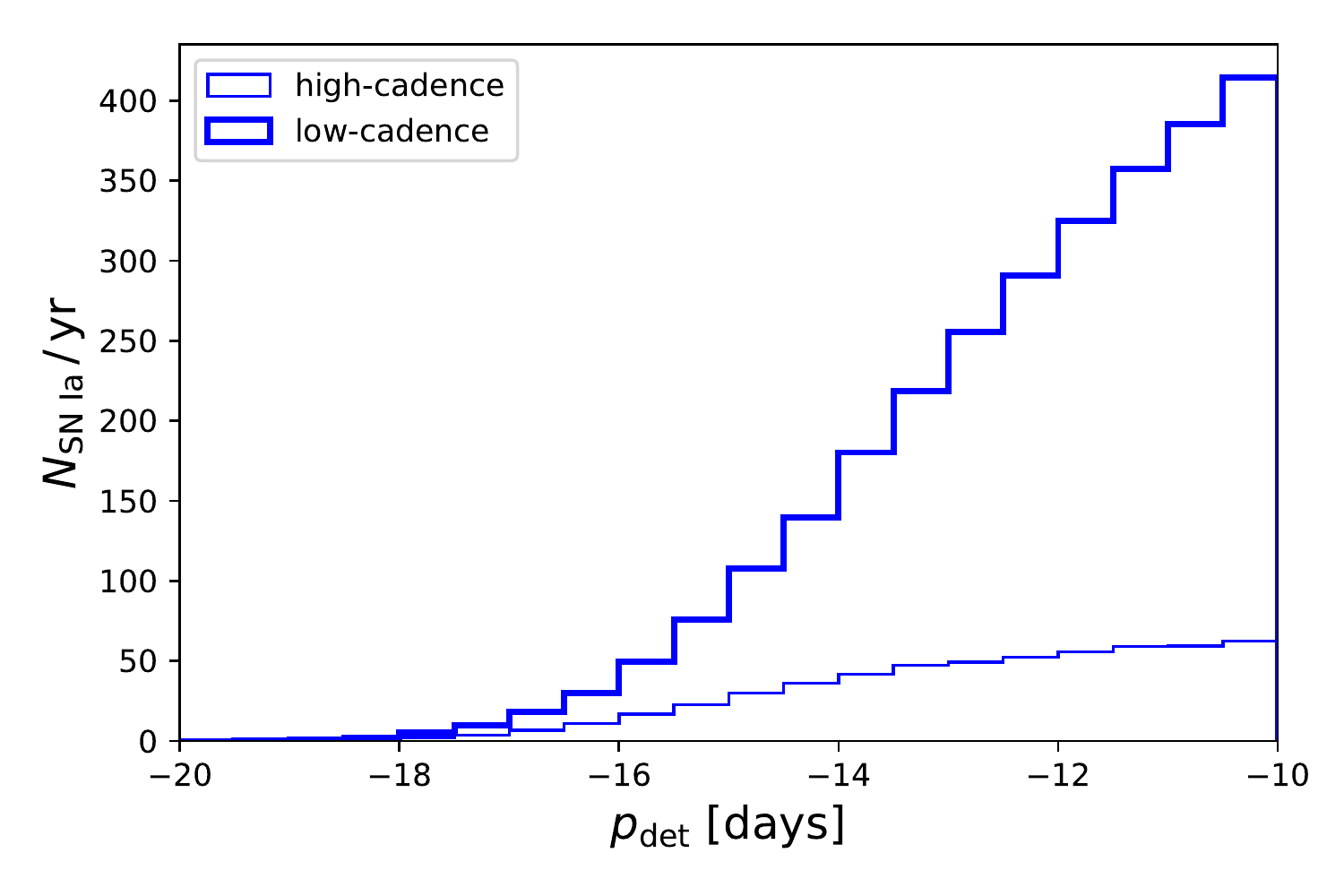}
  \includegraphics[width=0.49\textwidth]{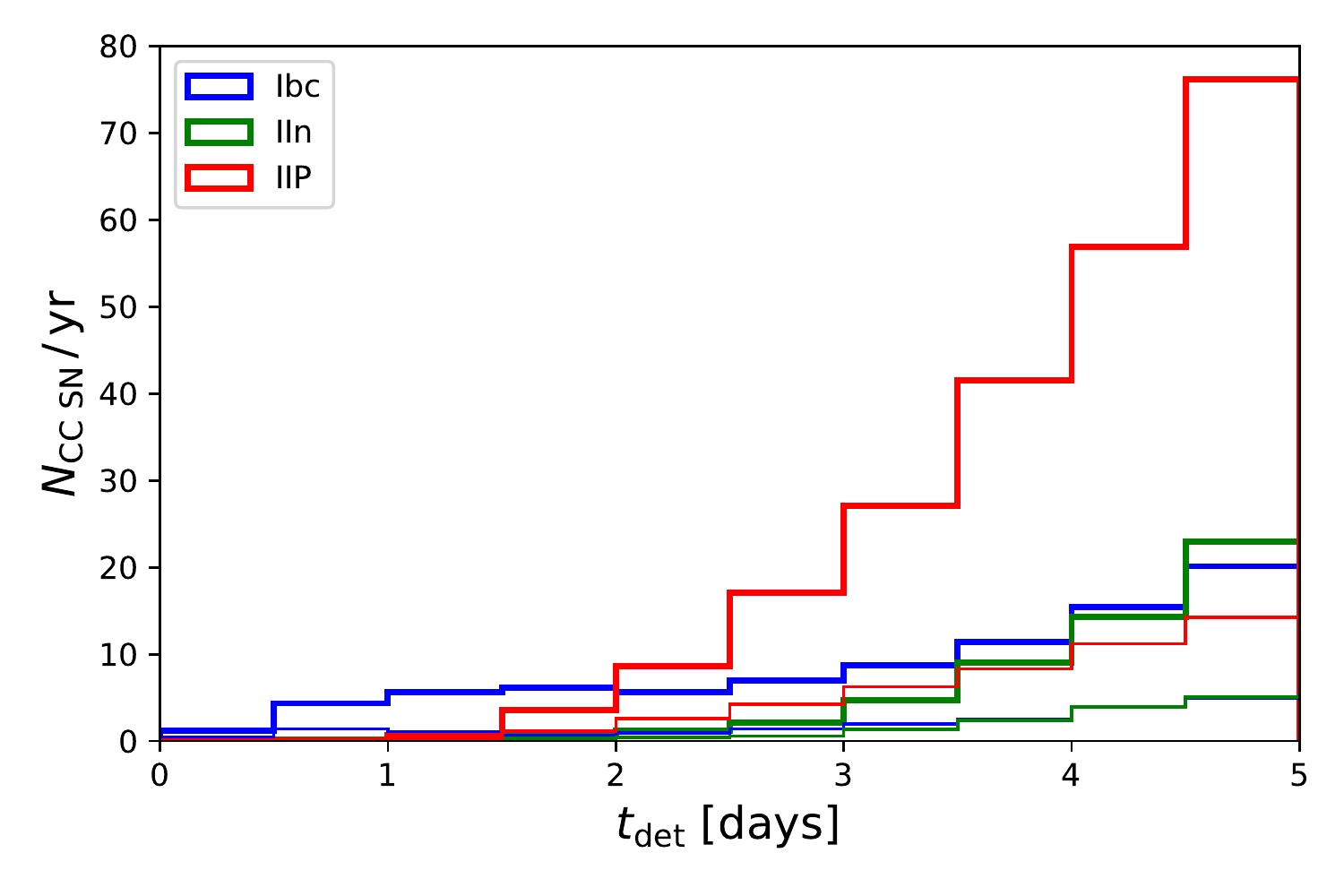}
  \caption{\emph{Left:} Discovery phases $p_{\rm det}$ w.r.t.\ $B$-band peak for the
    simulated SNe~Ia. The thick line shows the SNe in the public and $i$-band 
    (``low-cadence'') surveys while thin line shows only
    the ones found in high-cadence fields. \emph{Right:} Time of discovery
    $t_{\rm det}$ w.r.t.\ explosion for CC~SNe. Thick lines again show SNe in 
    low-cadence surveys while thin lines show high-cadence fields.}
  \label{fig:hist-p-det}
\end{figure}

Given the large survey speed of ZTF, many transients will be discovered very
shortly after explosion, some potentially in the same night. Many early
discoveries are expected to be made using the public survey alone but the
high-cadence part of the partnership survey will allow even earlier detections
and give tighter constraints on the time of explosion because the coordinates of
the transient will have been observed during the previous night (instead of
three nights ago) or potentially even earlier the same night.

\begin{table}[tp]
	\caption{Expected numbers per year of SNe discovered early for
          varying detection phases $p_{\rm det}$ w.r.t.\ $B$-band peak
          for SNe~Ia or times of detecion $t_{\rm det}$
          w.r.t.\ explosion for CC~SNe found either in the
          high-cadence survey or in the low-cadence parts of the 
          extra-galactic survey, i.e the public and $i$-band surveys. 
          (The different notations are due to the differences
          in definition of $t_0$, i.e.\ the time of peak or the time
          of explosion, respectively.)}
  \vspace{0.2cm}
	\label{tab:n-p-det}     
	\centering                
	\begin{tabular}{lcccc}
		\multicolumn{1}{c}{\bf SNe~Ia} & $p_{\rm det}<-15$ & $p_{\rm det}<-17$ & $p_{\rm det}<-18$ & $p_{\rm det}<-19$\\
		\hline                   
		high-cadence & 65 & 7.84 & 1.78 & 0.11\\
         low-cadence & 191.36 & 18.1 & 3.2 & 0.19 \\
                \hline\hline
		& $t_{\rm det}<5$ & $t_{\rm det}<3$ & $t_{\rm det}<2$ & $t_{\rm det}<1$  \\
		\hline
    \multicolumn{1}{c}{\bf SNe~Ib/c} &&&&\\
    %\hline
		high-cadence & 19.28 & 5.96 & 3.62 & 1.85 \\
         low-cadence & 66.42 & 24.02 & 13.69 & 3.65 \\
		\hline
    \multicolumn{1}{c}{\bf SNe~IIn} &&&&\\
    %\hline
		high-cadence & 14.04 & 1.29 & 0.34 & 0.06 \\
         low-cadence & 41.49 & 3.17 & 0.79 & 0.1 \\
		\hline
    \multicolumn{1}{c}{\bf SNe~IIP} &&&&\\
    %\hline
		high-cadence & 48.21 & 8.22 & 1.43 & 0.02 \\
         low-cadence & 183.53 & 21.8 & 2.91 & 0.02 \\
	\end{tabular}
\end{table}

To determine the phase at which phase a simulated supernova
was detected, we recorded the time of the second $5\sigma$-detection
of the SN and subtracted the parameter $t_0$ for the transient model
used. For SALT2 this is the time of $B$-band maximum but the Nugent
models of the CC~SNe this is the time of explosion. Thus, we show the
distributions for SNe~Ia and CC~SNe in separate panels of
Fig.~\ref{fig:hist-p-det}. The histograms were limited to five days
after explosion for the CC~SNe and to phases $p<-10$~days for
SNe~Ia. The plots show discoveries from the high-cadence survey
compared to the rest of the survey.  Since the high-cadence survey
covers an area about an order of magnitude smaller than the public
survey, we expect to find fewer transient in total in this limited
area. However, we expect the discoveries to be earlier, on average, for a
higher cadence. Accordingly we find that the median phase of SN~Ia
discovery relative the $B$-band maximum is -8.4~days in the high-cadence 
survey opposed to -5.6~days for the whole survey (including partnership 
time). The expected numbers of early discoveries per year is shown in
Table~\ref{tab:n-p-det}. We expect to find about five SNe Ia at phases
earlier -18~days and about 250 at -15~days. Of the CC~SNe only
SNe~Ib/c appear likely to be found within a day (about five or six per
year). A similar number of SNe~IIP may be discovered within two days
but only one SN~IIn is expected.

While the statistics discussed so far show that ZTF will produce
lightcurves of transients starting very close to explosion, we have
not yet addressed the issue how determine that these transients are
potentially young. The most important metric for that is the time
since the last pointing to the transient's coordinates, for which it
was not detected. This information was extracted for the simulated
lightcurves along with the phases of detection as the interval
between the first of the required two detections at $5\sigma$-level
and the previous lightcurve point that was below the threshold and
therefore would in reality only be an upper limit. In a more realistic
setting, we would also need to account for the varying depth of the
observation and only count limits that are enough to exclude something
as bright as the newly discovered transient. Since we assumed a
constant depth for simplicity, we do not require that level of
precision in our approach.

However, the time since the last non-detection in and of itself is
insufficient to find early transients. Many SNe that are too distant to
be detected early will also cross the detection threshold between two
nights or during the same night. The simulations show for example that
800 SNe~Ib/c will be found with a non-detection in the previous night
or later while only 5.5 are expected to be found in the within a day
of explosion. To weed out distant SNe, a redshift cut needs to be
applied as well. For this we will assume that the host galaxy redshift
can always be obtained from catalogs. Additionally the use of catalogs
will allow us to filter out known active galactic nuclei, which are
another possible source of contamination of the transient sample. In
reality this will not be possible in all cases because the host galaxy
cannot always be identified and redshift catalogs are not
complete. Therefore only a certain fraction of the simulated SNe can
be discovered this way. To estimate this fraction, we can use the findings 
of ZTF's predecessor surveys, PTF and iPTF. Of the SNe (CC and Ia) found 
by them 25\% could be associated with a galaxy that has a spectroscopic 
redshift in the SDSS catalogs and 78\% have a host galaxy with a photometric 
redshift \cite{host-paper-hangard}. Therefore, this efficacy of these 
criteria will mostly rely on photometric redshifts.

\begin{table}[tp]
	\caption{Expected number per year of SNe observed early ($p_{\rm det} <
    -17$~days for SNe~Ia and $t_{\rm det} < 2$~days for CC~SNe) and/or with a
    non-detection the night before or during the same night ($\Delta t_{\rm
      non-det} < 1.5$~days).}
  \vspace{0.2cm}
	\label{tab:n-dt-det}     
	\centering                
	\begin{tabular}{l|ccc|ccc}
    & \multicolumn{3}{c|}{$z < 0.03$} & \multicolumn{3}{c}{$z < 0.05$} \\
    & early \& & only & only & early \& & only & only \\
    & non-det. & early & non-det. & non-det. & early & non-det. \\
    \hline
    \multicolumn{1}{c|}{\bf SNe~Ia} &&&&&& \\
    high-cadence &  2.13 &  2.53 &  4.03 &   4.9 &  5.52 & 19.81 \\
    low-cadence  &  1.65 &  9.79 &  7.33 &  3.16 & 15.98 & 35.65 \\
    \hline
    \multicolumn{1}{c|}{\bf SNe~Ib/c} &&&&&& \\
    high-cadence &  1.42 &  1.73 &  3.56 &  2.28 &  2.78 & 16.04 \\
    low-cadence  &  1.30 &  7.18 &  6.81 &  2.06 &  11.0 & 31.25 \\
    \hline
    \multicolumn{1}{c|}{\bf SNe~IIn} &&&&&& \\
    high-cadence &  0.24 &  0.28 &  0.94 &   0.3 &  0.34 &  4.71 \\
    low-cadence  &  0.15 &  0.77 &  1.64 &  0.15 &  0.79 &  8.02 \\ 
    \hline
    \multicolumn{1}{c|}{\bf SNe~IIP} &&&&&& \\
    high-cadence &  1.21 &  1.29 & 17.32 &  1.32 &  1.42 & 73.35 \\
    low-cadence  &  0.54 &  2.79 & 32.44 &  0.56 &  2.91 & 132.24 \\
    \end{tabular}
\end{table}

Using these criteria for selection the advantages of the high-cadence
survey for finding early transients become much more obvious. For
instance, about 89\% of the SNe~Ia at redshift $z<0.05$ that are found
at a phase $p < -17$~days in the high-cadence survey also have an
observation without a detection during the same night or the night
before (see Table~\ref{tab:n-dt-det}). While basing the selection only
on low redshift and the time since last non-detection fails to filter
out all SNe that are found later than the selected phase, this number
is only about three times larger (14.91 later-than-specified
discoveries per year compared to 4.9 per year discovery before the
selected phase cut). For the rest of the extra-galactic survey (public
and $i$-band surveys), on the other hand, only about 20\% of the early
SNe can be found this way.

\subsection{Spectroscopically complete survey of bright transients}
\label{sec:spectr-surv}

For studies of the populations of transients in the local universe it
is essential to build catalogs of spectroscopically confirmed SNe to
fixed magnitude limit. This will help us to account for biases in the
higher-redshift parts of our data (e.g.\ for cosmological constraints)
and the redshift completeness of local galaxy catalogs (see
\cite{2017arXiv171004223K}).

\begin{figure}[tp]
  \centering
  \includegraphics[width=0.49\textwidth]{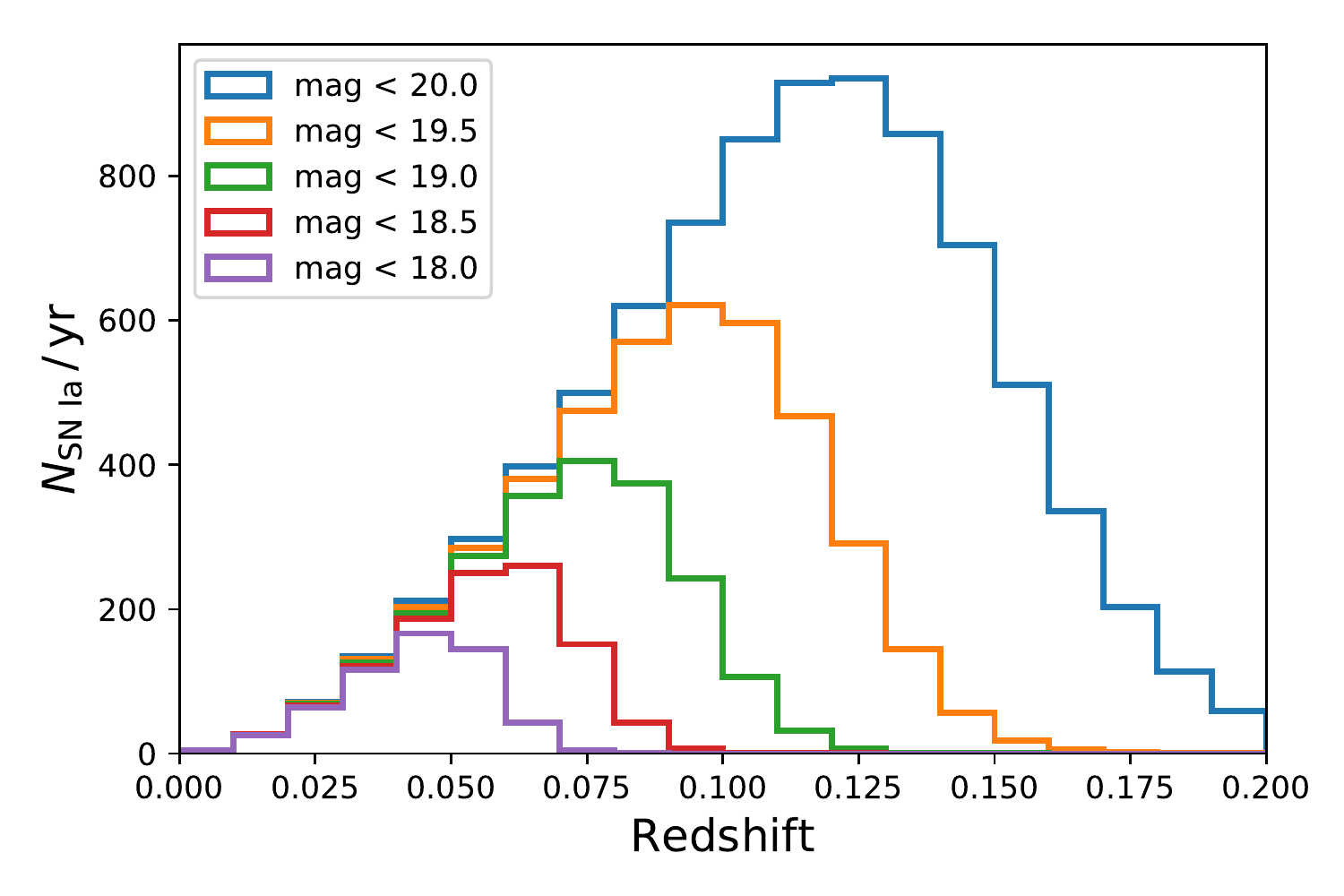}
  \includegraphics[width=0.49\textwidth]{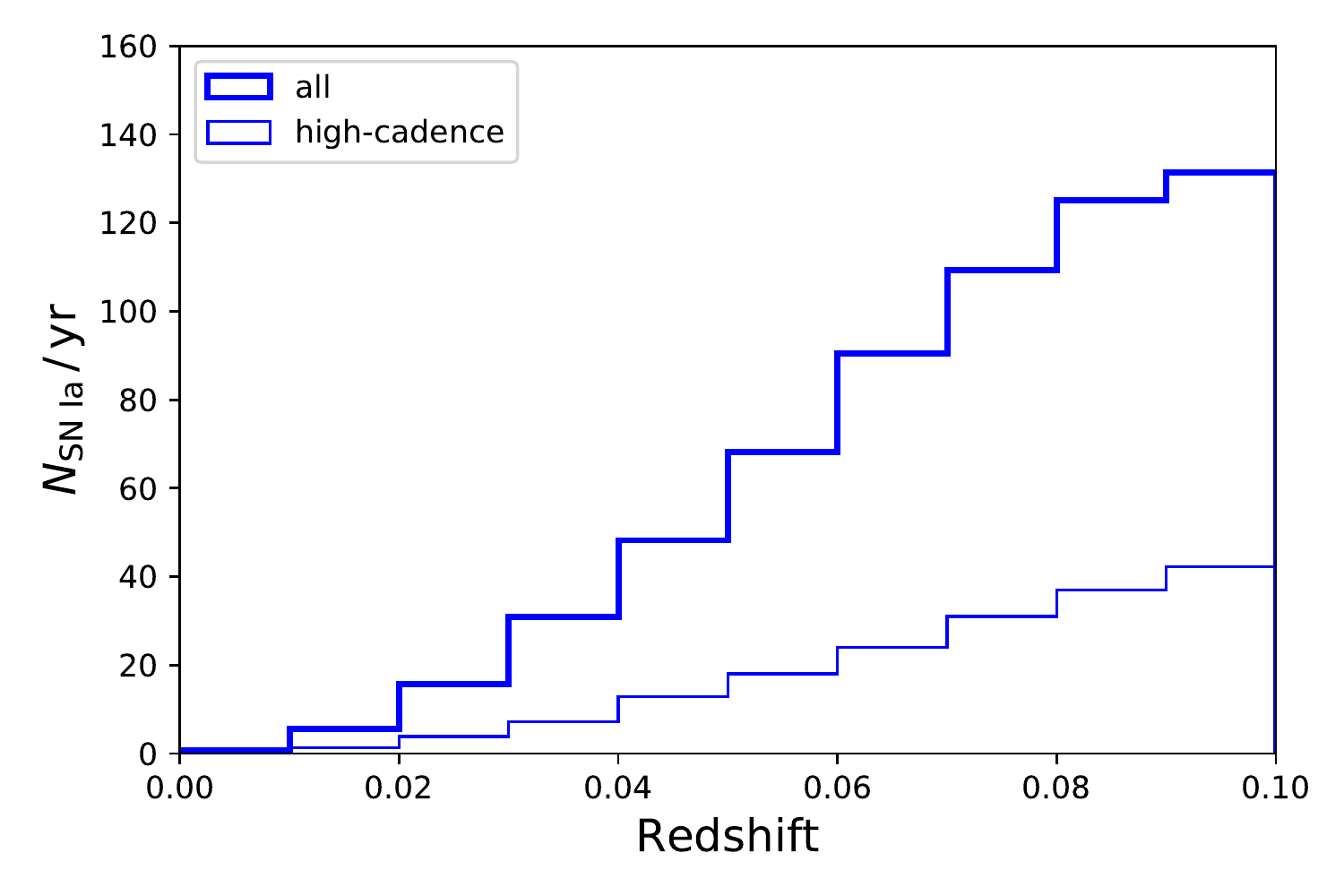}
  \caption{\emph{Left:} Redshift distribution of SNe~Ia for various peak magnitude
    cuts. \emph{Right:} Redshift distribution of SNe~Ia that pass the cosmology
    criteria.}
  \label{fig:hist-z-ia}
\end{figure}

In addition to the wide-field camera on P48, the ZTF collaboration has
access to time with the Spectral Energy Distribution Machine (SEDM;
\cite{2018PASP..130c5003B}), an integral field spectrograph mounted on
the Palomar 60-inch (P60) telescope. This will allow the spectroscopic
classification of all transients found by ZTF to a peak brightness of
18.5~mag and potentially fainter.
%Based on our simulations (see Table~\ref{tab:n-sne-maglim}), we
%expect 908~SNe brighter than 18~mag and about as many (860) between
%that and 18.5~mag.
Based on our simulations (see Table~\ref{tab:n-sne-maglim}), we expect
1768~SNe brighter than 18.5~mag (with $\sim$~50\% of those brighter than
18~mag) and about as many (1702) between that and 19~mag.  We find
that such a survey will probe the population of SNe out to a redshift
of 0.1 (in case of SNe Ia, see left panel of
Figure~\ref{fig:hist-z-ia}) with median redshifts around 0.05, see
Table~\ref{tab:n-sne-maglim}.

% [TODO: Add a sentence or two on how many spectra we might be able to get?]

% The exact
% magnitude limit up to this kind of survey can be performed will depend on the
% exact performance of the spectrograph.

\begin{table}[tp]
	\caption{Expected number per year of SNe peaking brighter than a given
    magnitude in both $g$- and $r$-band and their median redshifts $z_{\rm
      med}$. Note that these numbers are based on the brightest magnitudes in
    the simulated lightcurve (typically within one or two day of peak
    brightness) and not on the peak brightness of the model used for the
    simulation.}
  \vspace{0.2cm}
	\label{tab:n-sne-maglim}     
	\centering                
	\begin{tabular}{l|rr|rr|rr|rr|rr}
    & \multicolumn{2}{c|}{$\textrm{mag} < 18$} &
      \multicolumn{2}{c|}{$\textrm{mag} < 18.5$} &
      \multicolumn{2}{c|}{$\textrm{mag} < 19$} &
      \multicolumn{2}{c|}{$\textrm{mag} < 19.5$} &
      \multicolumn{2}{c}{$\textrm{mag} < 20$}  \\
    & $N$ & $z_{\rm med}$ & $N$ & $z_{\rm med}$ & $N$ & $z_{\rm med}$ & $N$ & $z_{\rm med}$ 
    & $N$ & $z_{\rm med}$ \\
    \hline
    %Ia  &   568 & 0.045 &  1116 & 0.056 &  2218 & 0.071 &  4352 & 0.090 &  8500 & 0.114 \\
    %Ibc &   115 & 0.040 &   216 & 0.049 &   405 & 0.061 &   747 & 0.074 &  1373 & 0.088 \\
    %IIn &    52 & 0.038 &   103 & 0.048 &   205 & 0.061 &   410 & 0.077 &   840 & 0.099 \\
    %IIP &   191 & 0.023 &   366 & 0.029 &   703 & 0.036 &  1334 & 0.044 &  2557 & 0.054 \\
    Ia  &   568 & 0.045 &  1116 & 0.056 &  2218 & 0.071 &  4352 & 0.090 &  8500 & 0.114 \\
    Ibc &   115 & 0.040 &   216 & 0.049 &   405 & 0.061 &   747 & 0.074 &  1373 & 0.088 \\
    IIn &    52 & 0.038 &   103 & 0.048 &   205 & 0.061 &   409 & 0.077 &   840 & 0.099 \\
    IIP &   173 & 0.023 &   333 & 0.029 &   642 & 0.036 &  1222 & 0.045 &  2339 & 0.055 \\
	\end{tabular}
\end{table}

\subsection{SN Ia lightcurves for cosmology}
\label{sec:results-ia-cosmo}

For cosmology, we are interested in obtaining SNe Ia lightcurves that provide
precise measurements of the distance modulus. As SNe Ia are not perfect standard
candles but require an amount of standardization based on their width and color
\cite{1993ApJ...413L.105P}, we can only use lightcurves that are well
sampled around peak in all three bands. Thus, all SNe in this sample are in the
fields of the $i$-band program. Additionally we restrict this dataset to SNe at
redshift $z<0.1$ because we will not have a flux-complete sample beyond that
redshift and therefore a cosmological analysis would be affected by Malmquist bias. The
full set of criteria is as follows:
\begin{itemize}
\item The redshift is less than 0.1.
\item The first detection is at least 10 days prior to $t_0t$,
  i.e.\ the time $B$-band maximum.
\item The last observed epoch is at 30 days after maximum.
\item There are at least three $i$-band pointings between 10 days prior to and 15
  days after maximum.
\end{itemize}

Based on 100 simulations using the SALT2 model, we find a median number of
626 SNe Ia matching these criteria. Their redshift and peak magnitude
distribution are shown in the right panel of Figure~\ref{fig:hist-z-ia}. The
median redshift is 0.075 and the median peak brightness is 18.5~mag in $g$-band.
Thus only half of these SNe will be covered when obtaining a spectroscopically
complete samples described in section~\ref{sec:spectr-surv}. The remaining
SNe~Ia will thus have to be classified using other resources. As ZTF will run
for three years, we can expect to collect $\sim1800$ lightcurves of the quality
described above. The cosmological implications of such a dataset will be
discussed in the next section.

\section{Discussion}
\label{sec:discussion}

In the previous section we have shown how ZTF will provide both many early
discoveries of transients and large datasets of the most common type of SNe.
The former will greatly help us understand the physics of relativistic explosions
and supernovae but a full discussion of this would be beyond the scope of this
paper. The latter, on the other hand, can be summarized by some cases in
which the sample of SN~Ia lightcurves described in
section~\ref{sec:results-ia-cosmo} will benefit cosmological studies.

\subsection{Dark energy constraints}
\label{sec:w0-wa}

\begin{figure}[tp]
  \centering
  \includegraphics[width=0.49\textwidth]{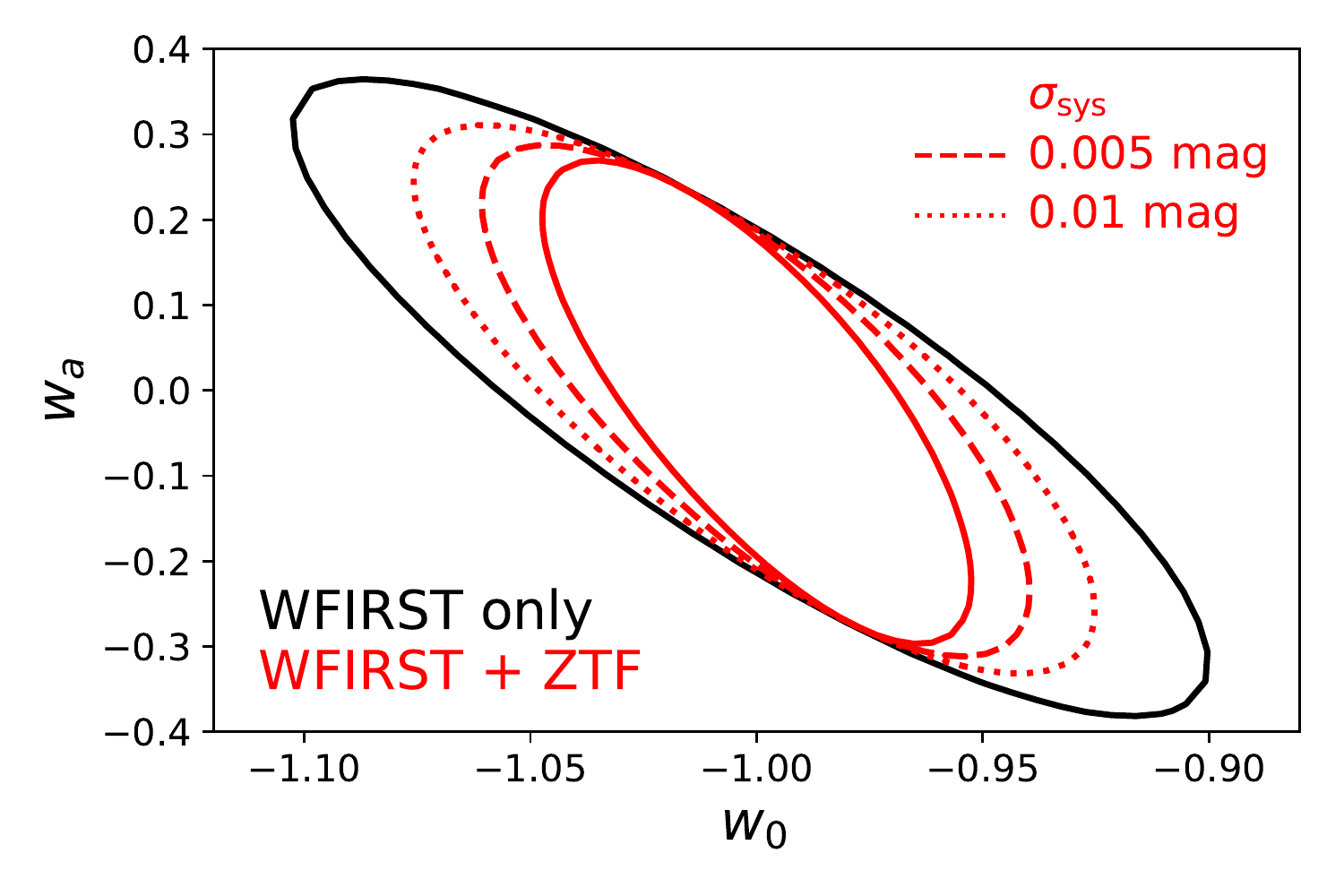}
  \includegraphics[width=0.49\textwidth]{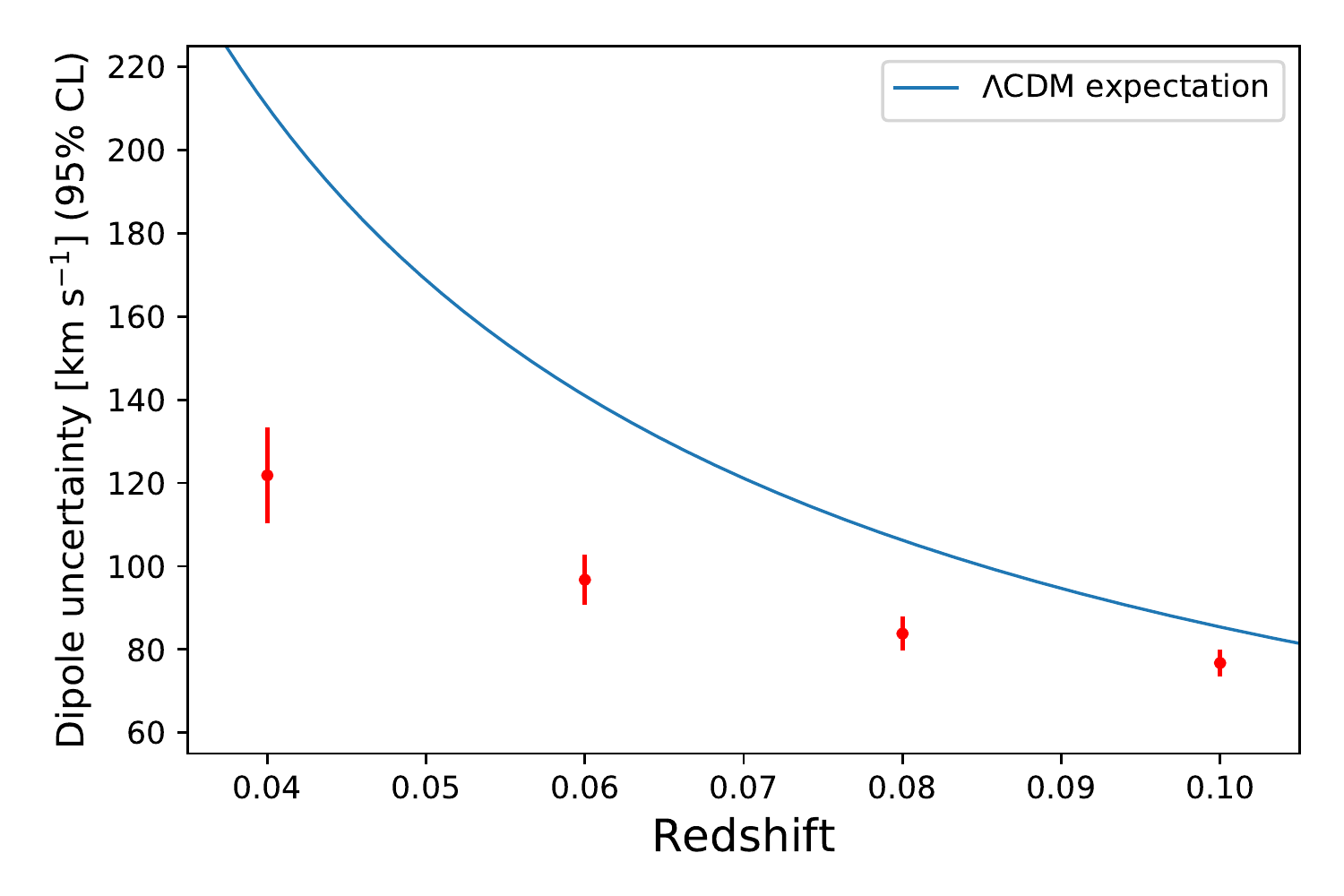}
  \caption{\emph{Left:} Expected confidence regions ($68.3\%$ CL) of
    the dark energy e.o.s.\ parameters $w_0$ and $w_a$ for a flat
    $w_0w_a$CDM cosmology model for WFIRST with and without the
    addition of a ZTF low-redshift SN~Ia data set. The solid red line
    assumes that the datasets are both perfectly calibrated while the
    other red lines show the effect of a systematic uncertainty
    $\sigma_{\rm sys}$ of 0.005~mag (dashed line) or 0.01~mag (dotted
    line). \emph{Right:} Expected uncertainties for a peculiar
    velocity dipole derived from three years of ZTF SN~Ia data as a
    function of the maximum redshift used in the fit. The error bars
    correspond to the nMAD of all possible combinations of three out
    of the 100 simulations that were run. The solid line shows the
    expected size of a dipole from $\Lambda$CDM structure formation,
    showing the uncertainty required to determine whether the measured
    dipole matches the expectation for an isotropic universe.}
  \label{fig:cont-w0wa}
\end{figure}

A large, low-redshift data set of SNe Ia will greatly improve the
constraints on the dark energy equation of state (e.o.s.) by anchoring
the Hubble diagram.  This can best be demonstrated by comparing how
the constraints from a future survey of higher-redshift SNe~Ia, e.g.\
by the Wide-Field InfraRed Survey Telescope (WFIRST), improve when a
low-redshift sample is added to the data set. We used the distance
modulus uncertainties and systematic covariances, binned in redshift,
from the simulations presented in \cite{2017arXiv170201747H}. We
removed the bins at redshifts $z < 0.1$ because these bins are based
on simulations of the Foundation survey \cite{2018MNRAS.475..193F} 
and here we wish to investigate the impact that ZTF can have on its own.
%% We used the equations in Appendix C.1 of the final report by the
%% Science Definition Team and WFIRST Project \cite{2013arXiv1305.5422S} to
%% calculate the binned distance modulus uncertainties that we can expect from
%% WFIRST. For this we read the numbers of SNe per bin from Figure~C-1 in the
%% report by eye.
For ZTF we use the equations in Appendix C.1 of the final report by the
Science Definition Team and WFIRST Project
\cite{2013arXiv1305.5422S} to determine the binned uncertainties we
expect from ZTF,
\begin{equation}
  \label{eq:de-eos-unc}
  \sigma_{\rm stat}=\sqrt{\frac{(\sigma_{\rm meas})^2+(\sigma_{\rm
      int})^2+(\sigma_{\rm lens})^2}{N_{\rm SN}}},
\end{equation}
where $\sigma_{\rm meas} = 0.08$~mag, $\sigma_{\rm int}=0.08$~mag,
$\sigma_{\rm lens}=0.07\cdot z$~mag, and the values for $N_{\rm SN}$ are
binned numbers of SNe~Ia shown in the right panel of
Figure~\ref{fig:hist-z-ia}, multiplied by a factor three for the
expected duration of the survey.
%% Furthermore we use the binned
%% distance moduli for the JLA data set~\cite{2014A&A...568A..22B}
%% because it would be unrealistic do estimate the impact of the ZTF
%% sample without acknowledging that there already is low-redshift SN~Ia
%% data.

As a metric of the quality of the dark energy constraints, we
calculate the confidence regions for the parameters $w_0$ and $w_a$ of
a flat $w_0w_a$CDM cosmology model
\cite{2001IJMPD..10..213C,2003PhRvL..90i1301L}, i.e.\ a model where
the dark energy e.o.s.\ evolves with redshift and is parameterized as
\begin{equation}
  \label{eq:de-eos}
  w = w_0 + w_a \frac{z}{1+z}.
\end{equation}
This method is similar to the method defined by the Dark Energy Task
Force \cite{2006astro.ph..9591A}, using the inverse area of the
confidence region as a figure of merit (FoM). As a fiducial model we
assume $w_0=-1$ and $w_a=0$, which corresponds to a flat $\Lambda$CDM
cosmology. To constrain the estimate further we include a simple
Gaussian prior on the matter density parameter $\Omega_{\rm M}$ based
on the most recent results from the Planck Collaboration
\cite{2018arXiv180706209P}, setting its width to 0.007.  As the binned
uncertainties from our ZTF simulations do not include an estimate of
the systematic uncertainties of the distance estimates (e.g.\ from
calibration uncertainties or evolution of SN~Ia brightness with
redshift), we included a offset term that shifts the ZTF data relative
to WFIRST as a nuisance parameter in the likelihood. For the prior on
this offset we assumed two cases, both Gaussian, one with a width of
0.005~mag and the other with 0.01~mag. The latter value is our target
for the calibration precision, which is the main contributor to the
systematic uncertainties in SN~Ia cosmology, while the other is a more
optimistic scenario.

The resulting contours are shown in
Figure~\ref{fig:cont-w0wa}. Including ZTF data in such an analysis
will greatly improve the constraints, increasing the figure of merit
by 30\% compared to WFIRST alone, given a systematic uncertainty
for ZTF of 0.01~mag (73\% for 0.005~mag).

\subsection{Local anisotropy}
\label{sec:aniso}

An addition to constraining dark energy, the data set that ZTF
will collect will allow more precise measurement of structure and
anisotropy in the local universe. One possible test of anisotropy is by looking
for a peculiar velocity dipole or bulk flow in the low-redshift data. If this
dipole exceeds the expectation from $\Lambda$CDM structure formation this could
be indicative of effects beyond the standard model of cosmology.

To estimate the uncertainties of a dipole estimate using ZTF data, we apply the
same method as in \cite{2013A&A...560A..90F}, where the dipole formula from
\cite{2006PhRvL..96s1302B} was used to determine the effect of peculiar
velocities on the luminosity distance:
\begin{equation}
  \label{eq:bonvin-dipole}
  \tilde{d}_L(z,\vec{n},\vec{v}_d)=d_L(z)+\frac{(1+z)^2}{H(z)}\vec{n}\cdot\vec{v}_d,
\end{equation}
where $d_L(z)$ is the unperturbed luminosity distance, $\vec{n}$ is the radial
unit vector corresponding to the SN's coordinates and $\vec{v}_d$ is the bulk
flow velocity vector. The velocity vector could then be determined by
statistically inference such as minimizing the following $\chi^2$ expression:
\begin{equation}
  \label{eq:chisq-dipole}  
  \chi^2=\sum\limits_{i}\frac{\left|\mu_i-5\log_{10}\left(\left(d_L(z_i)+\frac{(1+z_i)^2}{H(z_i)}\vec{n}\cdot\vec{v}_d\right)/\left(10 \mathrm{pc}\right)\right)\right|^2}{\sigma_i^2},
\end{equation}
where $\mu_i$ is the measured distance modulus of a SN Ia and
$\sigma_i$ is its uncertainty, for which we use the same formula as in
the previous section (equation \ref{eq:de-eos-unc}). For this
expression the resulting covariance matrix of the velocity dipole is
then expressed as
\begin{equation}
  \label{eq:cov-dipole}
  \mathrm{Cov}(\vec{v}_d)=\left(
    \sum\limits_i\left(\frac{5(1+z_i)^2}{\ln10\tilde{d}_L(z_i)H(z_i)\sigma_i} \right)^2
    \vec{n}_i\vec{n}_i^T\right)^{-1}.
\end{equation}
In order to sample how the differences in redshift distribution and sky coverage
between the 100 one-year simulation affect the dipole as it would be determined
by a three-year survey, we calculated this covariance matrix for all 161700
possible combinations of three out of the 100 simulations, using several
redshift cutoffs for the data to see the evolution of the uncertainty with the
radius of the sphere, in which the bulk flow is determined. Since the
expression in equation~(\ref{eq:chisq-dipole}) is not completely linear in
velocity, the uncertainty depends on the best-fit value. Thus we chose to assume
a velocity dipole in the direction of the Shapley supercluster, which roughly
corresponds to the direction most studies of the bulk flow find. For the
amplitude we assumed that the velocity decreases with the radius of the sphere
as $v\sim250\left( \frac{100 \mathrm{Mpc}} {d}\right)\mathrm{km~s^{-1}}$, where
$d$ is the comoving distance from the observer. This is the approximate
expectation for a dipole arising from random fluctuations in $\Lambda$CDM (see
e.g.\ \cite{2010ApJ...712L..81K}). The right panel of Fig.~\ref{fig:cont-w0wa}
shows the uncertainty of the bulk flow for the selection redshift cutoffs as
well as the expected value from structure formation. This shows that the ZTF
SN~Ia data set will allow us to constrain the bulk flow at the level of what we
expect to find due to structure formation.

\section{Comparison to first ZTF detections}
\label{sec:comp-ztf}

\begin{figure}[tp]
  \centering
  \includegraphics[width=0.49\textwidth]{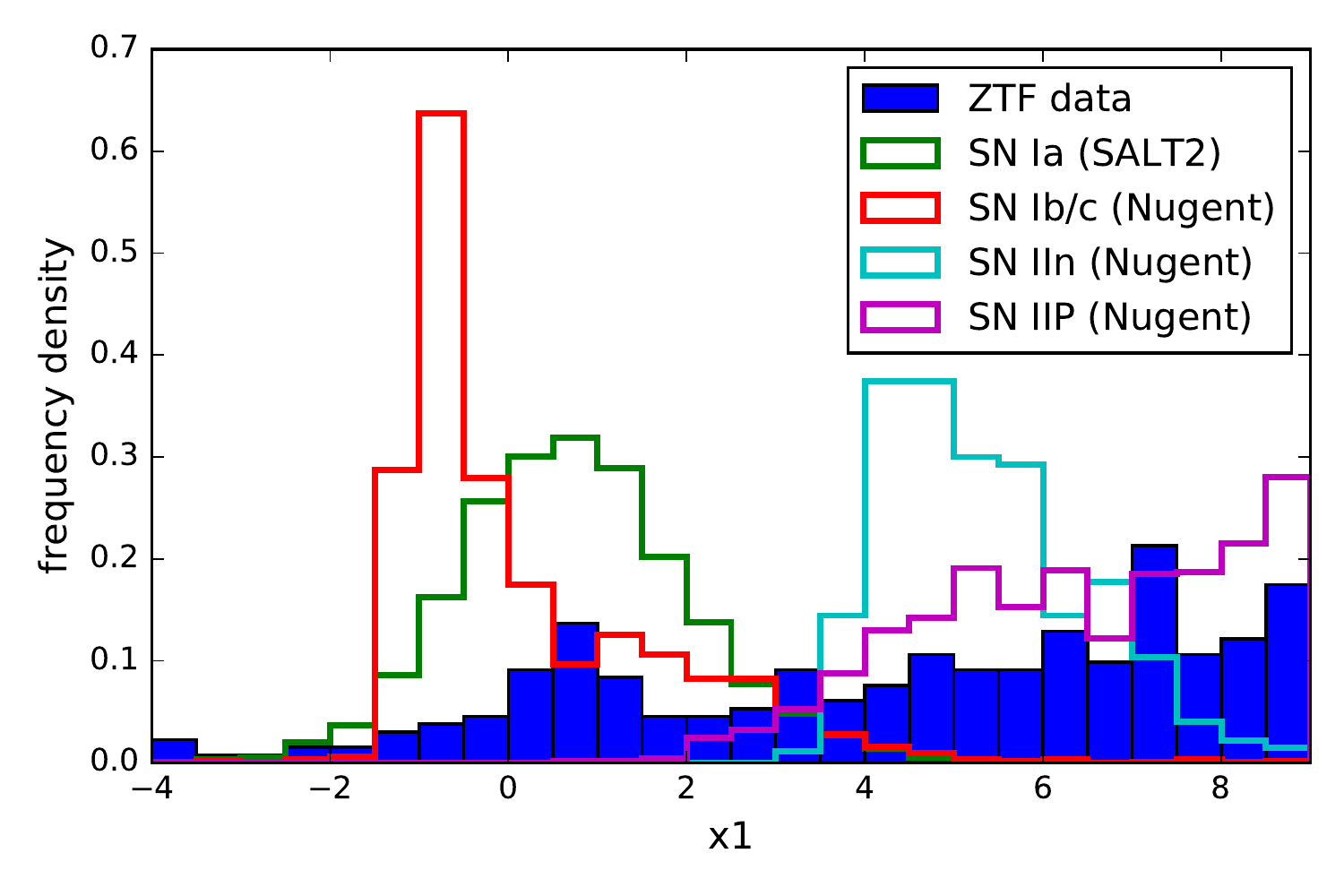}
  \includegraphics[width=0.49\textwidth]{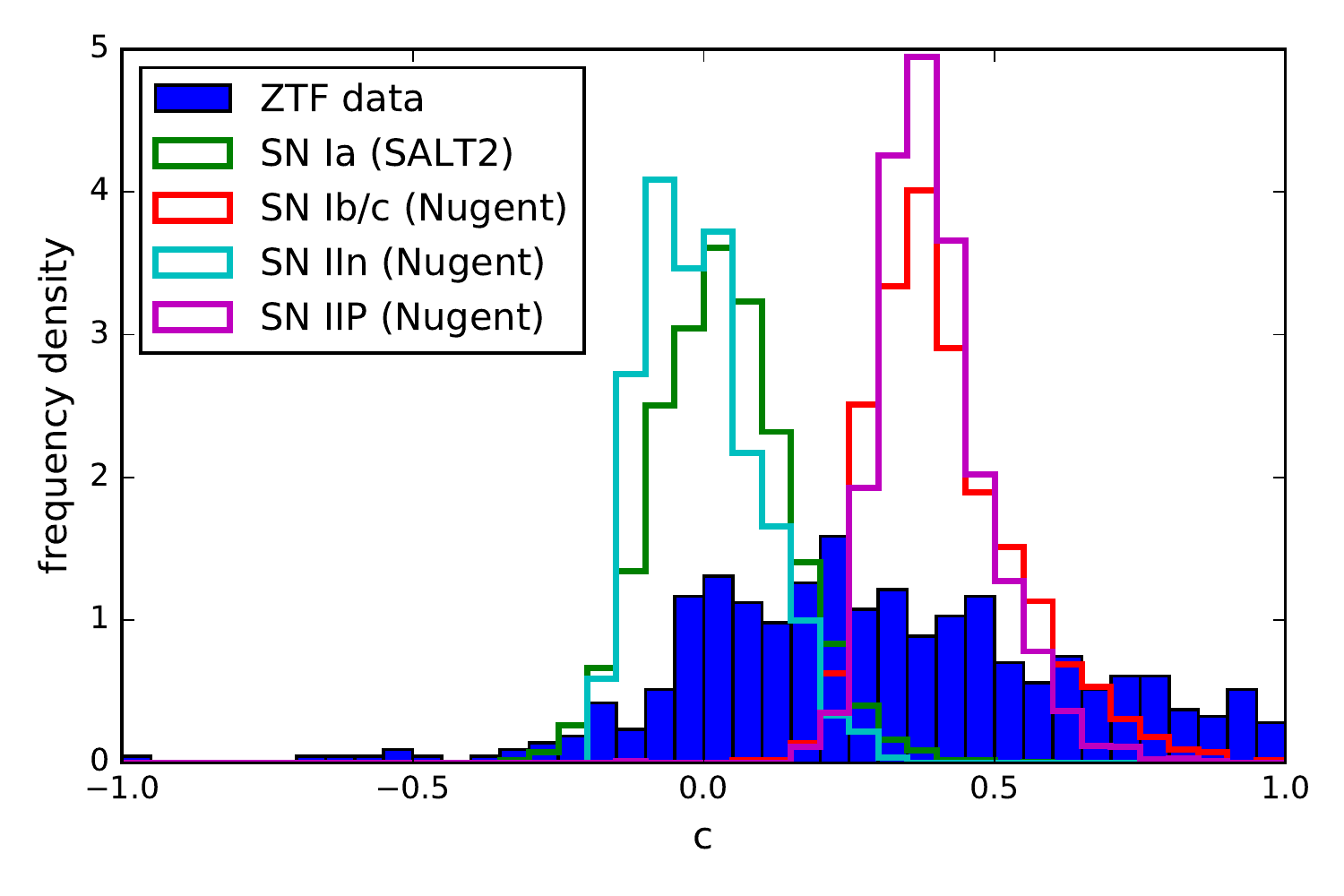}
  \caption{Histograms of the SALT2
    ``stretch'' parameter $x_1$ (left panel) and color parameter $c$
    (right panel) based on lightcurve fits to ZTF public alert data
    for selected fields and detection in July 2018 (blue solid bars)
    and simulated lightcurves based on SN models restricted to the
    same fields and time of detection. }
  \label{fig:comp-ztf-sim}
\end{figure}

Since ZTF started observing in early 2018, many transients have
already been found and these discoveries can thus be compared to the
predictions from the simulations in this paper. However, at this time
a full measurement of supernova rates would be beyond the scope of
this paper. We will therefore limit this section to a quick comparison
of how many SNe were found in a specific set of fields in July. These
fields were chosen because that had fully-built reference images by
the beginning of July and were observed until the end of the
month. The ZTF discoveries were extracted from the public alert stream
using the AMPEL broker \cite{ampel-nordin}. Since we used no ZTF 
partnership data in this analysis, we also removed the partnership 
epochs from simulated lightcurves to match the cadence. For both the 
alerts and the simulated lightcurves we required that there were at 
least five points brighter than 20~mag during the month of July.

For each selected lightcurve, we extracted basic parameters by fitting
the SALT2 model to it. We further included the redshift as a fit
parameter because a host redshift cannot be obtained automatically
without a greater effort that would be beyond the scope of this
section. While this model is intended for standardizing SNe~Ia
brightnesses for cosmological distance measurements, its fit results
can also be used to distinguish SN types. The ``stretch'' parameter
$x_1$ and the color parameter $c$ are most useful for this because
different types of SNe have different rise times and color indices and
thus SNe of types other than Ia will not have a distribution centered
on 0 in one of these parameters or both. The absolute peak magnitude
of a transient could also be used to distinguish SN types but the
photometric redshifts obtained from the fit are too inaccurate to give
a useful distribution. Based on the distributions of the lightcurve
parameters for the simulated SNe (see Figure \ref{fig:comp-ztf-sim},
we find that transients with $-2<x_1<2$ and $-0.2<c<0.2$ are most
likely to be SNe Ia. While a selection based on these criteria is not
expected to yield a complete or pure sample, it is a sufficient for a
simple comparison. Based on these criteria we find 37 transients among
the filtered ZTF alerts, which is in good agreement with the 44 SNe
(almost exclusively of type Ia) expected based on the simulated data
sets. We have further checked whether these object were in fact
SNe~Ia, finding that 19 of them have been spectroscopically typed as
SNe~Ia while no spectra were taken of the rest.

\section{Conclusion}
\label{sec:concl}

We have presented a software package, \texttt{simsurvey}, for the
simulation of lightcurves for photometric transient surveys. The
package can be used for most types of extra-galactic transients as
long as the user provides a spectral time series, from which the
photometry can be synthesized at various redshifts. Additionally the
package contains built-in simulators for common supernova types such as
SNe~Ia. Furthermore the code is not specific to any particular survey
and can be used to simulate lightcurves based on any schedule and
telescope configuration.

As an example of its utility, we have simulated the lightcurves of
type~Ia supernovae and several types of core-collapse supernovae as
they are expected to be observed by the Zwicky Transient Facility. The
simulations included both the wide public survey as well as
partnership surveys focusing on smaller areas but using a higher
cadence or an additional filter. Our significant findings include:
\begin{itemize}
\item ZTF will find thousands of SNe per year, about 1800 of which will peak at
  a magnitude < 18.5~mag.
\item ZTF will be able to find on the order of 10 very young SNe per year and will be able
  to identify them as young through a combination of host redshifts (from
  catalogs) and the time since the last observation of its coordinates without a
  detection (mostly driving by high-cadence observations of a limited set of
  fields).
\item Within a year about 600 SNe~Ia with sufficient lightcurve quality for
  cosmology will be found. A data set built up over the course of three years
  will contain about an order of magnitude more low-redshift supernovae than
  current samples and will be essential for future studies of the dark
  energy equation of state as well as studies of the local anisotropy.
\end{itemize}

\noindent Lastly, we have compared our simulations to part of the discoveries
made so far by ZTF based on the public alert stream for a month and a
limited number of fields that had reference images by the start of
July. Based on this comparison we found a good agreement between our
simulations and the first discoveries of the survey.

\acknowledgments

We would like to thank Rahul Biswas for stimulating discussions and
comments and Dan Scolnic for providing us with WFIRST simulation
results. Funding from the Swedish Research Council, the Swedish Space
Board and the K\&A Wallenberg foundation made this research possible.
This project has received funding from the European Research Council
(ERC) under the European Union's Horizon 2020 research and innovation
programme (grant agreement n$^\circ$759194 - USNAC). Based on observations
obtained with the Samuel Oschin Telescope 48-inch and the 60-inch
Telescope at the Palomar Observatory as part of the Zwicky Transient
Facility project. ZTF is supported by the National Science Foundation
under Grant No. AST-1440341 and a collaboration including Caltech,
IPAC, the Weizmann Institute for Science, the Oskar Klein Center at
Stockholm University, the University of Maryland, the University of
Washington, Deutsches Elektronen-Synchrotron and Humboldt University,
Los Alamos National Laboratories, the TANGO Consortium of Taiwan, the
University of Wisconsin at Milwaukee, and Lawrence Berkeley National
Laboratories. Operations are conducted by COO, IPAC, and UW.

\paragraph{Facilities}
PO:1.2m

\paragraph{Software}
\texttt{Astropy} \cite{2018AJ....156..123T}, 
\texttt{Numpy} \cite{numpy:2011}, 
\texttt{Matplotlib} \cite{Hunter:2007},
\texttt{Scipy} \cite{scipy:2001},
\texttt{SNCosmo} \cite{2016ascl.soft11017B}

\bibliographystyle{JHEP}
\bibliography{references}{}

\end{document}